\def\parder #1;#2;{{\partial #1\over \partial #2}}   
\def\les {\ {\buildrel < \over {_{_{\sim}}}}\ }
\def\ges{\ {\buildrel > \over {_{_{\sim}}}}\ }                           
  
\def\eps@scaling{1.0}%
\newcommand\epsscale[1]{\gdef\eps@scaling{#1}}%
\newcommand\plotone[1]{%
 \centering 
 \leavevmode 
 \includegraphics[width={\eps@scaling\columnwidth}]{#1}%
}%
\newcommand\plottwo[2]{%
 \centering 
 \leavevmode 
 \columnwidth=.45\columnwidth 
 \includegraphics[width={\eps@scaling\columnwidth}]{#1}%
 \hfil 
 \includegraphics[width={\eps@scaling\columnwidth}]{#2}%
}%
\newcommand\plotfiddle[7]{%
 \centering 
 \leavevmode 
 \vbox\@to#2{\rule{\z@}{#2}}%
 \includegraphics[%
  scale=#4, 
  angle=#3, 
  origin=c 
 ]{#1}%
}%

\newcommand\etal{{\it et al.\ }}

\newcommand \prl{Phys. Rev. Lett}

\newcommand \no{$n^0$}

 \documentclass [12 pt, epsf]{article}

                             \usepackage{color}  
\usepackage{graphicx}  
                                                  
\begin{document}
\begin{center}

\section*{SN1987A--a Testing Ground for the KARMEN Anomaly} 
\subsection*{I. Goldman$^1$, R. Mohapatra$^2$ and S. Nussinov$^1$}

\subsubsection*{
$^1$School of Physics and Astronomy, Tel Aviv University,
Tel Aviv, 69978, Israel}
\subsubsection*{
$^2$Department of Physics, University of Maryland, College Park, MD 20742, U.S.}
\end{center}

\abstract
 We show, that SN1987A can serve as an astrophysical  laboratory for
 testing the viability of the assertion  that a
new  massive neutral fermion is implied by 
the KARMEN data. We show that a wide range of the parameters
characterizing the proposed particle is ruled out by the above
constraints making this interpretation very unlikely.

\section{Introduction}

The KARMEN collaboration, which studies neutrino from pion decay at the
Rutherford laboratory has reported an anomaly in the time structure of
their signal, which deviates from that expected from the standard
decay into muons \cite{KAR1}, \cite{KAR2}. One possible interpretation of this, offered
in the experimental papers, is the production of a heavy neutral fermion
with a mass of approximately 33.9 MeV and a specific range of values for
the life times. It is the goal of this paper to study
the (astrophysical) viability of such a fermion using supernova 1987A data. 

 As is well known, 
the KARMEN detector is located $17.5$ m away from the target
in the ISIS spallation neutron source. Two pulses of $800$ MeV protons, of
$100{\rm  s}$ duration, and $0.32{\rm \mu s}$ separation, impinge every
$20$ milliseconds on the target \cite{KAR1}, \cite{KAR2}. The time distribution of $\sim
5\times 10^3$ events , where energy in excess of 10 MeV has been
deposited in the detector, fits very well the assumption that these
are interactions of neutrinos from stopped $\pi^+$ ( and subsequent
$\mu^+$) decays with a moderate, time independent background.
However, the time bin between $3.1$ to $4.1$ microseconds has an excess
of $\sim 100$ events over of the $ 550$ expected from the overall
fit.\cite{Oeler}

If this $4.5\sigma$ effect is not an accidental fluctuation or an
instrumental artifact, its explanation requires a slow ( $\beta\sim
\frac {1}{60}$) relatively monochromatic ``messenger'' traversing the
distance $L=17.5{\rm m}$ in a time of $L/(\beta c)\sim3.6\mu s$. It should
then deposit an energy $E\geq 10{\rm MeV}$ in the detector.
Since many (about $10^{24}$ ) neutrons were produced over the time, is
it possible to explain the anomaly via a neutron messenger?
The order of $10$ meter iron shielding 
corresponds to  $ n l \sigma\sim 60$ mean free paths of neutron interactions. One
would then expect the neutrons to be absorbed/ diffused and in any
event $\it{not}$ to come in a well defined time.

However, the transport of neutrons from target to detector can be
dominated by {\it one specific} crack or external path involving
reflections.  Even in this case the ``neutron $=$
messenger'' hypothesis may not be viable. Indeed, 

(i) The neutrons emerge from the target with high ( $\sim 100{\rm
  MeV}$) energies. There is no obvious , early moderator which is
  required to slow the neutrons down to  $\sim 100{\rm KeV}$
  energies. Further, there is no obvious mechanism for guaranteeing the 
  required time delay 
  \begin{equation}
 \frac{L_{path}}{c}\left<\frac{1}{\beta}\right>\sim 3.6\mu {\rm s}
\end{equation}
with angular brackets indicating average (over many collisions with
possible moderation) along the crack/external path length,  $L_{path}$.

(ii) The interactions in the anomalous time-bins are distributed
uniformly over the detector rather than near the front edge or the
termination of a putative crack.

(iii) The neutron shielding around the detector was enhanced in the
second phase of the experiment. This did not stop the steady buildup
of the excess of events in the anomalous bin.

(iv) The strongest argument against the neutron hypothesis comes
from the energy distribution of the events. Barring  the unlikely
possibility that  the detector had substantial amounts of $^3{\rm He}$
and/or $ ^{235} U$, the generic $(n,\gamma)$reactions of the slow
neutrons {\it  cannot} deposit  energy in excess of the $8$ MeV
nuclear binding. If the observed time-anomaly is enhanced/weakened
by an energy cut $E\geq 15{\rm MeV}$
, the case against a neutron
messenger will be considerably strengthened/weakened. Unfortunately, 
because of possible excess of events in the first energy bin $10\div
15$ MeV in figure 4 of \cite{Oeler} one cannot make this assertion yet.

For the rest of our discussion we will assume that the
KARMEN anomaly has no simple explanation. We thus have to address the 
bold hypothesis \cite{KAR1,Oeler} that a new neutral fermion \no\,
exists.
The properties of this, \no, (the ``Karmino'' implied by the KARMEN data) 
are:

(i)Its mass is precisely tuned to be

\begin{equation}
  m_{ n^0}=m_{\pi^+}- m_{\mu^+}- 5 {\rm KeV}= 33.906 \pm 0.005{\rm MeV}
\end{equation}
We assume that in a (small) fraction $(= B_r (\pi^+\rightarrow n^0+\mu^+))$ of
the cases, the stopped $\pi^+$ decays into \no$ + \mu^+$. The
monochromatic \no, tuned to have $\beta=1/60$, will then arrive at the 
KARMEN detector $ 3.6 \mu s$ later.

If some of \no\, particles 
decay via \no$\rightarrow e^+e^-\nu$ or via \no$\rightarrow
\gamma \nu$ 
while traversing the detector,  an energy $\geq
17{\rm MeV}$ (or exactly $17$ MeV, respectively) will be deposited. This would explain
the anomaly if the $B_r(\pi^+\to \mu^+$\no)\, and the decay rate say
$\Gamma_{n^0}\to e^+e^-\nu$  satisfy 

(ii)

\begin{equation}
 B_r(\pi^+\to \mu^+n^0) \Gamma(n^0 \to
e^+e^-\nu)=\frac{B_r(\pi^+\to \mu^+n^0}{\tau(n^0 \to
  e^+e^-\nu)}=2.6\times 10^{-11}{\rm s^{-1}}
\end{equation}

Is such a particle defined by equations  (2), and (3) above (and most
likely 
having week interactions in matter) consistent with other
terrestrial and astrophysical data? This issue has been
considered before \cite{barger}.  Recent new experimental
bounds \cite{PSI}, and new astrophysical considerations presented below 
motivate us  to
reconsider it.

\section{Terrestrial data and particle physics considerations}

The particle physics constraints on the properties of the singlet fermion
implied by the Karmen anomaly (Karmino) have been extensively studied in
several recent
papers \cite{barger,sar,others}. Here we revisit them to set the stage for
our ensuing discussion and also to incorporate the constraints from the
PSI search for the Karmino\cite{PSI}. We also find further constraints
on the parameters of Karmino within the context of some plausible
theoretical assumptions.

The striking kinematics of the  $\pi^+\to \mu^+n^0$  decay at rest, are
equally striking in
decays in flight. Energetic pions yield decay muons moving in
the same direction and with the same speed. This feature has been used in a 
recent experiment at PSI--which was inspired by the KARMEN 
anomaly --to obtain the remarkable bound \cite{PSI}

\begin{equation}
  B_r(\pi^+\to \mu^++n^0)\leq 2.6 \times 10^{-8}\  .
\end{equation}

Equation (3) then implies that

\begin{equation}
  \tau_{(n^0 \to e^+ e^-\nu)}\les 10^3 s\ .
\end{equation}

The lack of evidence for sharp $\sim 17$ MeV energy deposition for events 
in the anomalous bin, and  a theoretical bias (the need to use loops
rather than tree diagram) suggest that the above decay, rather than
$n^0\to \nu+\gamma$, dominates.

It is useful to parameterize the  $\pi\to \mu^+n^0 $ and
$n^0\to e^+e^-\nu $  processes via effective
local four-Fermi interactions. 

\begin{equation}
 \widetilde{G}_{\mu}\bar{\Psi}_d(x)\gamma _5\Psi_u(x)\
 \Psi_{n^0}(x)\Gamma '\Psi_{\mu}(x)\ ,
\end{equation}

and 

\begin{equation}
  \tilde{G}_{n^0}\bar{\Psi}_{n^0}(x)\Gamma''\Psi_{\nu_{d}}(x)\
  \Psi_e(x)\Gamma'\Psi_{e}(x)\ ,
\end{equation}
with Lorentz/flavor structure  which may be different from those for 
neutrinos in the standard Electroweak model. Still we can compare using 
just phase space considerations\footnote{For more detailed discussion
that takes into account the general Lorentz structure carefully, see
\cite{bob}.}, the expected ratio of the
decay rates: $\Gamma_n/\Gamma_{\mu}\approx \left(m_{n^0}/m_{\mu}\right)^5 
 \widetilde{G}_{\mu}^2/ G_F^2$
Using  $m_{n^0}/m_{\mu}\approx 1/3, \ \Gamma_{\mu}\approx 4\times 10^5 s^{-1}$
and equation (5) we find that

\begin{equation}
\frac{\widetilde{G}_{n^0}^2}{G_F^2}\ges 0.7\times 10^{-6}\ .
\end{equation}

The local effective Lagrangian implies also the crossed scattering
process (at energies $E_{\nu}>>m_{n^0}^2/2m_e\sim {\rm GeV}$)

\begin{equation}
\bar{\nu}_d + e^-\to e^- + n^0\ ,
\end{equation}
with $\nu_d$ the neutrino appearing in the \no decay $ n^0\to e^+e^-\nu_d $,
and  the process $e^+e^-\to n^0
\bar{\nu}_d$ with
 crossection

\begin{equation}
\frac{\sigma(e^+e^-\to  n^0\bar{\nu}_d) }{\sigma(e^+e^-\to
  \nu_i\bar{\nu}_i)}\sim \frac{\widetilde{G}_{n^0}^2}{G_F^2}\   .
\end{equation}
This last process 
features in SN1987A and supernovae in general. We therefore need to know
the constraints on the coupling $\widetilde{G}_{n^0}$ from laboratory
experiments to study the atsrophysical viability of the Karmino.

The specific Lorenz structure of the standard $\pi^+\to \mu^+\nu$ (or
$\pi^+\to n^0\mu^+$) decay only
mildly affects the decay rate. Hence, 

\begin{equation}
\frac{\tilde{G}_{\mu}^2}{G_F^2} \beta\approx
   \frac{\Gamma (\pi^+\to n^0\mu^+)}{\Gamma(\pi^+)}\equiv B_r(\pi^+\to
  n^0\mu^+)\leq 2.6\times 10^{-8}\  ,
\end{equation}
where we included the phase  space suppression factor $\beta\sim1/60$ for
$ \pi^+\to n^0\mu^+$. From the last equation we find that

\begin{equation}
\frac{\widetilde{G}_{\mu}^2}{ G_F^2}\les 1.5\times 10^{-6}\ .
\end{equation}

Very short partial life-time in the $10^{-6}\rm {s}\leq\tau(n^0\to
e^+e^-\nu_d)\leq 0.6\times 10^{-3}{\rm s}$ range can be excluded by
the following ``theoretical arguments'' showing that no new
interactions stronger than the ordinary ``weak'' interactions are
allowed. Indeed, $\widetilde{G}_{n^0}=G_F \sqrt{\epsilon_{n^0}}$ implies
that 

\begin{equation}
  \tau(n^0\to
e^+e^-\nu_d)=\tau(\mu) \left(\frac{m_{\mu}}{m_{n^0}}\right)^5
\epsilon_{n^0}^{-1}=0.7\times 10^{-3}\epsilon_{n^0}^{-1} {\rm s}\ .
\end{equation}
From this we see that life-times shorter than $10^{-3}$ seconds would
require values  of $\widetilde{G}_{n^0}$ larger than $G_F$.
However, in the presently accepted approach to field theories one {\it 
  cannot} postulate at will new non-renormalizable four-Fermi
interactions. The latter should be viewed only as the low energy
limit of massive boson exchanges. The bilinear vertices of the
boson exchange tree-diagram, $\bar{\Psi}B\Psi$, which are parts of the
interaction Lagrangian in a fully renormalizable and acceptable field
theory.
To date , the only such known bosons are the vectorial (spin $1$)
gluons, photon, and $W^+,\ W^-,\  Z^0$ which generate the strong,
electromagnetic and weak interactions , respectively. Yet, the
exchanges could also be scalars generating $aS +bP$  four vertices
instead of the (V-A) combination for $W_{\mu}$.

The effective $\widetilde{G}_{n^0}$ is then of the form $\widetilde{g}_1
\widetilde{g}_2 /m_{\widetilde{X}^2}
\propto \widetilde {G}_{n^0}$,
with $\widetilde{g}_1\widetilde{g}_2$v representing the $ \widetilde{X} e^+e^-$ 
  and $\widetilde{X}n^0 \nu_d$ vertices and $m_{\widetilde{X}}$ is the mass
  of the new boson. This should be compared with $g_W^2/m_W^2\propto
  G_F$  (with the same proportionality constant). 
Since, $g_W\approx 0.7$ and we do not wish to entertain
$\widetilde{g}>1$, we need $m_X\leq m_W$ in order to allow for
$\widetilde{G}> G_F$.
However, we will now have either a charged $X^-$ particle exchange in $ 
e^- \nu_d\to X^- \to e^-n^0$ or (and) a neutral $X^0$ in the crossed
channel.: $e^+e^-\to X^0\to \bar{\nu}_dn^0$, with $ m_{X^-}$ or (and)
$m_{X^0}$ smaller than $m_W$.

The new LEP runs at $W_{CM}\geq 200{\rm\, GeV}$ strongly exclude 
new charged particles of mass smaller $85{\rm GeV}$. Also, the new
$X^0$ would have strongly manifested as a new narrow resonance in
electron-positron scattering in the Tristan and LEP regimes. (We note
also that for electron-positron scattering at energies larger than
the mass of $X$ the latter could still manifest if there is an
emission of an initial internal bremsshtralung photon.)

The LEP bounds ($N_{\nu}\les 3.01$ ) on sequential, $I_{weak}\neq 0$,
particles with masses $\leq 40{\rm GeV} $ imply
that  $n^0$  is  a weak-Iso-spin singlet. 
If the decays $ \pi^+\to \mu^+
n^0$ and/or $n^0\to e+e^-\nu_d$ do
involve also left -handed quarks or leptons then by using $SU(2)_W$
rotation 
the processes could be related to their isospin analogues. Thus $\pi^+\to
\mu^+n^0$
would be related to $\pi^0\to \bar{\nu}n^0$ which is  controlled by the
same $\tilde{G}_{\mu}$ . The
latter may again manifest via $\nu_{\mu}+N\to N'+ n^0$ in the hot
proto neutron star. The latter process is more effective there (has a
smaller Boltzman factor suppression) than $\mu^+n\to p +n^0$. Note,
 however that the charged current process is 
 guaranteed by the very existence of four-Fermi coupling,
equation (6), with no additional assumptions.

The $\pi^+\to \mu^+n^0$
process suggests that,  if muonic lepton number is conserved then  \no 
 has $L_{\mu}=1$ and
hence the decay neutrino $\nu_d$ in \no $\to e^+e^-\nu_d$ should
also have $L_{\mu}=1$. If we
make the natural (though not mandatory!) assumption that
$\nu_d=\bar{\nu}_{\mu}$, then by
crossing and isospin rotation, we could generate yet another decay
mode for the muon.
\begin{equation}
  \mu^+\to e^+\nu_e n^0\ .
\end{equation}
Apart from a small ($\sim 40\%$) phase space correction due to
                                $m_{n^0}/m_{\mu^+}\sim 1/3$ the rate of
the new mode is given just like for $\Gamma(\mu\to e \nu\nu)$ by

\begin{equation}
  \Gamma(\mu^+\to e^+\nu_e n^0)\approx\frac{\bar{G}_{\mu}^2
    m_{\mu}^5}{192 \pi^3}\ ,
\end{equation}

hence

\begin{equation}
\frac{ \Gamma(\mu^+\to e^+\nu_e n^0) }{\Gamma(\mu^+\to e^+\nu\nu)}\approx
\frac{\tilde{G}_{\mu}^2}{G_F^2}\equiv \epsilon_{\mu}\ .
\end{equation}
Since the muon decay rate and the electron spectrum and polarization
in this decay have been studied carefully over the last four decades,
one could deduce that :

\begin{equation}
  \epsilon_{\mu}\leq 10^{-3}\left(
{\rm if} \ \ \ \  n^0\to e^+e^-\nu_{\mu}\right)
\end{equation}
Using equation (8) this implies a  lower bound on
  $\tau_{n^0\to e^+e^-\nu_{\mu}}$:

\begin{equation}
  \tau_{n^0}\to e^+e^-\nu_{\mu}\geq 1  {\rm s}\ \ \ \
  \left({\rm  if}\ \ n^0\to
e^+e^-\nu_{\mu} \right)\  ,
\end{equation}
in addition to the model-Independent  upper bound $\tau_{n^0\to
e^+e^-\bar{\nu}_d}\leq 10^3\ {\rm s}$
 implied by the PSI upper
bound on $B_r(\pi\to \mu^+n^0)$.

To summarize, we note that the PSI experiments in conjunction
with some
plausible theoretical assumptions lead to bounds on the  lifetime of
\no\ :
 $1~s\leq \tau_{n^0} \leq 10^3~s$. Furthermore, the
strength of the four-Fermi coupling of \no\  to electrons defined by the
parameter $\epsilon_{n^0}$ (see the previous section) has a lower bound
$\epsilon_{n^0}\geq 0.7\times 10^{-6}$ which will imply lower bounds on
the production of \no in supernovae environments. We consider the impact
of this on SN1987A observations in the next section.

\section{Limits Derived from  Supernovae, Notably Supernova 1987A}

In the standard scenario for a type-II supernova \cite{araa}, most ($\sim
99.7\%$) of the
collapse energy is emitted via neutrinos. The latter are trapped in
the dense hot proto neutron star for  few seconds during which they
diffuse out. The optical signal is delayed by $\sim 1\div 3$ days,
after  the shock
emanating from the core traverses the progenitor's envelope.
The energy  of the shock ($\sim 10^{51}$ erg) manifest as the 
kinetic energy of the ejected envelope. Only about a percent i.e,
$10^{49}$ erg manifest in the (bolometric) electromagnetic signal.

This scenario-- largely worked out before the SN87A-- explosion has been
brilliantly verified there. A notable exception is the fact that the progenitor was
not a red supergiant but a blue supergiant. Accordingly, the optical
signal appeared  an usually short time of $\sim 3$ hours after the
neutrino pulse that marked the core collapse.

Many putative novel, features in particle physics can modify the
above scenario and/or leave other enduring traces. Hence,
considerations of SN1987A data (and supernovae in general ) strongly limited
\cite {Seckel-Raffelt}
 the interactions of axions, right handed neutrinos, 
(generated via $m_{{\rm Dirac}}$ and/or via precession of neutrino spin in magnetic
field when the neutrino posses an anomalous magnetic moment
\cite{GNAA}, \cite{barmoh}).
Finally electromagnetic neutrino decays are severely limited.

We would like to find the ranges of lifetime $ \tau$ and branching
$B_r$ allowed by the observations from SN1987A. For most part we will
only assume that $B_r/\tau\sim 2.6\times 10^{-11}{\rm s}^{-1}$
 as required by the magnitude of the KARMEN
anomaly. In  terms of the two four-Fermi couplings introduced above
this equivalent to

\begin{equation}
  \epsilon_{n^0}\epsilon_{\mu}\equiv
  \frac{\widetilde{G}_{n^0}^2}{G_F^2}
  \frac{\widetilde{G}_{\mu}^2}{G_F^2}\approx 10^{-12}\ .
\end{equation}
We note that $ B_r\approx 2.6 \times 10^{-8}$ ( the upper bound of the
recent PSI experiment) and
the corresponding $\tau\approx 10^3{\rm s}$ imply roughly equal
$\epsilon_{n^0},\ 
\epsilon_{\mu}$ ( actually $\epsilon_{n^0} =0.6 \times 10^{-6}\  {\rm
  and}\ \epsilon_{\mu}=1.5\times 10^{-6} $).
As will be shown below,
the existence of the \no\, s with such properties can have quite
dramatic effects on supernovae astrophysics.

\subsection{Neutrino emission}

The production of \no\, in the newly formed hot core can proceed via both the 
$\widetilde{G}_{n^0}$  and  $\widetilde{G}_{\mu}$ couplings. That is via 
\begin{equation}
 e^+e^-\to n^0+\nu_d, 
\end{equation}
with crossection
\begin{equation}
  \sigma\approx \tilde{G}_{n^0}^2 E_e^2\ ,
\end{equation}
and via  nuclear scattering 
\begin{equation}
 \nu N\to n^0N\ , 
\end{equation}
with crossection
\begin{equation}
  \sigma\approx \tilde{G}_{\mu}^2 E_{\nu}^2\ .
\end{equation}
The latter process can be viewed as due to $\pi^0$ exchange. At  low
($\leq 50$ MeV)
 energies this yields 
an effective local four-Fermi vertex.  If the
$\pi^0\to \nu n^0$ vertex {\it cannot} be inferred from $\pi^+\to \mu^+
n^0$ we have instead $\mu^+ n\to n^0 p$ with crossection

$$\sigma(  \mu^+ n\to n^0 p)\approx \tilde{G}_{\mu}^2 m_{\mu}^2 \ .\eqno(22')$$
These processes should be compared with the standard neutrino
scattering crossection, $\sigma_W\approx G_F^2 E_{\nu}^2$.
In the dense ($ \rho\ges 10^{15}{\rm gr cm}^{-3}$, i.e. super-nuclear
density) and hot ( $ T\ges 30{\rm MeV}$) core, the resulting neutrino mean
free path is: 

\begin{equation}
  l=\frac{1}{n\sigma_W}=100{\rm cm}.
\end{equation}
It implies a total number of neutrino collisions, during diffusion
\begin{equation}
  N_{\nu N collisions}\approx \left(\frac{R}{l}\right)^2\ges 10^8
 ,
\end{equation}
 where $R\approx 30{\rm km}$ is adopted for the radius of the very hot
 core. In a fraction 
$\epsilon_{\mu}f_B(m_{n^0}/T)$
of the collisions we could have, instead of the usual weak
scattering, \no\ production 
$  \nu_{\mu}+N\to n^0+ N\ ,$
or 
 $  \mu^+n\to p+n^0$.

Note that the Boltznman factor, $f_B$,  reflects the
energy distribution of the initial particles (neutrinos in the
present case). The \no\, themselves need not be in thermal
equilibrium. 
For $T\sim 30 $ MeV the appropriate   $ f_B\approx
\left(m/T\right)^{3/2}e^{-m_{\mu}/T}$ is $\sim1$ and for T$\sim 10$ MeV it is $\sim
0.5$. When only  $  \mu^+n\to p+n^0$  contributes, the corresponding Boltzman
factor is $f'_B\approx 2.7 \left(E/T\right)^{3/2} e^{-m_{\mu}/T}$ which 
  for $T=10{\rm MeV}$ is $\sim 1/4$.

When $ \epsilon_{n^0}\gg \epsilon_{\mu}$  \no\, production is dominated
  by $ e^+e^-\to n^0\nu_d$. To estimate the rate of
the latter, we note that all collapse models indicate a sizable
electron density in the proto-neutron star
($\frac{n_e}{n_N}\geq 0.2 $ \cite{borr}) during the first few seconds. The
number of weak interactions induced $ e^+e^-$ collisions per positron
during these few seconds is now given by:

\begin{equation}
  N_{Weak-collisions}\approx  \left(\frac{R}{l_W}\right)^2\approx 10^6 
  \   ,
\end{equation}
where we have used $l_W=\left(n_e \sigma_W\right)^{-1}\approx 1000{\rm cm}$.

Since 
$\frac{\sigma(\nu_{\mu}+N\to n^0+N)}{\sigma(\nu_{\mu}+N)}\approx  \frac 
{\widetilde{G}_{\mu}^2}{G_F^2}\approx\epsilon_{\mu}$, and 
$\frac{\sigma (e^+ e^-\to n^0 +\nu_d)}{\sigma (e^+e^-\to \nu \bar{\nu}}
\approx  \frac 
{\widetilde{G}_{n^0}^2}{G_F^2}\approx\epsilon_{n^0}$,
\no\, production occurs in a fraction of $ \epsilon_{\mu} f_B$  (or
$\epsilon_{n^0} f_B$) of the $10^8 \ \nu N$
 (or the $10^6$ weak $e^+e^- $ ) collisions. Altogether, the number of
produced \no s is 
$N_{total}(n^0)\approx N_{total}(\nu_i)  \left(10^8 \epsilon_{\mu} +
10^6 \epsilon_{n^0}\right) f_B
\geq  20 f_B N_{total}(\nu_i)\approx (20\div 5)  N_{total}(\nu_i)$,
In the penultimate step we used $x+y\geq 2 \sqrt{x y}$ and
$\epsilon_{\mu}\epsilon_{n^0}=10^{-12}$ , and finally substituted
$f_B=1$ (or $0.6$) for reactions {\it r$_1,\  r_2$} respectively and
$T=10{\rm MeV}$.
\footnote{
Due to the chemical potential in the supernova core there is a
suppression of the positron density in the supernova core, which is not
explicit in our discussion, since the effect cancels in
 $\frac{\sigma (e^+ e^-\to n^0 +\nu_d)}{\sigma (e^+e^-\to \nu \bar{\nu}}
\approx  \frac 
{\widetilde{G}_{n^0}^2}{G_F^2}\approx\epsilon_{n^0}$.
The suppression factor is $\sim \left(\frac{T}{\mu}\right)^3 e^{-\mu/ T}$.
Taking $\mu\sim 200$ MeV and $T\sim 60$ MeV, this factor is of order
$2\times 10^{-3}$. There is a more dominant mechanism for $n_0$ production
that comes from $\nu_{\mu} + e^- \rightarrow e^- + n_0$
which does not suffer from chemical potential suppression. As long as the
muon neutrinos are equally abundant in the core, this process adds to the 
 first term in  
$N_{total}(n^0)\approx N_{total}(\nu_i)  \left(10^8 \epsilon_{\mu} +
10^6 \epsilon_{n^0}\right) f_B
\geq  20 f_B N_{total}(\nu_i)\approx (20\div 5)  N_{total}(\nu_i)$,
 and leaves our conclusions unchanged.}

Note the the above rate of \no\, particle production (which already
tends to exceed by $5\div 20$ the initial number of parent neutrinos) 
is actually a minimum achieved for a particular choice of
$\epsilon_{\mu}/\epsilon_{n^0}\approx N_{\nu N\ {\rm collisions}}/N_{e^+e^-\ {\rm Weak\ 
  collisions}}\approx 100$. Generically equation (29)  above leads to $
N_{n^0}>f N_{\nu_i}$ with  $f\sim 10^2\div
10^6 $! \no particles \footnote{If only (22') rather than (20)
  operates then $10^6\epsilon_{n^0}\to 10^6 \epsilon_{n^0} 
(f_B'/f_B)$ but this decrease the total number of the \no\,
  particles by merely a factor 
  of 5.}.   What this apparently paradoxical result, really means is the
following. Our assumption that neutrinos continuously convert into
\no\, s during their entire ordinary, diffusion time 
$ t_{diff}\approx l_{mean free path} N_{coll} /c\sim 0.3{\rm s}$
is false. Rather, a sizeable fraction converts into \no on a much
shorter timescale $t_{convers}\approx t_{diff}/ f\ll 0.1{\rm s}$.

{\it  If} the \no\, particles escape  on this
short timescale, out of the hot proto-neutron star,
the momentary reaction balance  will  immediately shift toward the \no
s. Catastrophic cooling of the core on this short timescale will
then ensue. The resulting neutrino pulse would be drastically shortened
in time and reduced in intensity. The IMB and Kamiokande observation of 
SN1987A neutrino pulses lasted $5\div 10$ seconds and indicated that
the total energy emitted via neutrinos was $\ges 3\times 10^{53}$
erg, as expected. Hence, these observations by themselves
exclude almost the complete range of $B_r, \tau$ (or $\epsilon_{\mu},
\epsilon_{n^0}$)\, parameters and thereby the 
\no\, hypothesis itself.

To complete the argument let us estimate the escape time from the core 
for any single \no\, particle.
The \no\, particles diffuse out on  timescale of 
$t_{n^0, diff}<< t_{diff}/100\approx 3 {\rm ms}$,
once  the crossection of the \no\, s on ambient particles ($\sigma_{n^0
  N}, \sigma_{n^0 e}, \sigma_{n^0\nu}$) is ten
times smaller than the corresponding $\sigma_W$ (the neutrino or
other particles weak crossection). 
We will therefore make
the assumption:

 $A1$:
{\it The nuclear and other crossections  of \no particles are (at
least ten times) smaller than those of neutrinos.} 

This  assumption is clearly consistent 
with the value of $B_r$ and $\tau$(or $\widetilde{G}_{\mu}^2$ and
$\widetilde{G}_{\mu}^2$)  required  for the
KARMEN anomaly. The \no\, particles are devoid of color. Otherwise
they would Hadronize and certainly could not penetrate 10 meters of
iron. They are elecromagnetically neutral and as noted above are
also singlets of the full $SU(2) \times U(1)$ electroweak gauge group. Hence assumption 
A1 is eminently reasonable,  though given the meager information on
\no\, particles it is not however,  mandated in an absolutely model-independent manner.

In all the above discussions, we made the implicit assumption that
$\tau_{n^0}$ the  actual decay life-time of  \no is longer than
$R/c\sim 10^{-4}{\rm s}$. Therefore the bounds derived apply only for 
$\tau$ values exceeding $\sim 10^{-4}$.

Thus, \no with lifetime in the range $1~s\leq \tau_{n^0} \leq 10^3~s$
would be inconsistent with the SN1987A observations. A simple way to see
this is to note that the corresonding bounds for $\epsilon_{n^0}$ i.e.
$10^{-6} \leq \epsilon_{n^0} \leq 10^{-3}$ are in the range forbidden for
general coupling of weakly interacting particles to
electrons\cite{raffelt}. 

One way to avoid this bound would be to
give up the theoretically plausible assumption that \no\ has muonic
quantum
number and $\epsilon_{n^0}\sim 1$ in which case all the produced \no's get
trapped and this SN1987A constraint can be avoided. In this case, both the
\no\ and $\nu_d$ have to be sterile neutrinos. 
In this case also one can derive a constraint on $\epsilon_{n^0}$ from
big bang nucleosynthesis. The point is that if
$\epsilon_{n^0}$ is indeed of order one, it will inject sterile neutrinos
$\nu_d$ into the cosmic soup and they will count as one neutrino species.
Thus if the BBN constraint on additional neutrino species is much less
than one \cite{olive}, then the \no\, lifetime has to be less than $10^{-5}$
sec so that any population of $\nu_d$'s injected into the cosmic soup by
\no decays would be diluted by the QCD phase transition. This implies
roughly that $\epsilon_{n^0} \geq 0.5$. 

Another possibility is to postulate the existence of exotic interaction of
both the \no\ and the $\nu_d$ to quarks so that those interactions trap
the \no\'s in the supernova core. Clearly, such interactions must have
a very
specific property \cite{babu} i.e. the \no\ and $\nu_d$ must couple only
to the isosinglet current so that the $\pi^+$'s do not decay directly into
the \no\ in order to be cnsistent with laboratory observations i.e.
$\pi^+$
decay mostly to the $\mu^+n_0$. As shown in \cite{babu}, this also helps
to avoid BBN constraints. To be a bit more quantitative, we may note that
if the interaction is denoted by $
{\cal L}_{int}=\frac{G_F\epsilon}{\sqrt{2}}\bar{n^0}\Gamma\nu_d
(\bar{u}\Gamma u +\bar{d}\Gamma d)$,
then the \no-nucleon scattering cross section is given by
$\sigma_{n^0 N}\sim \frac{9}{\pi} (g^2_V+3 g^2_A)G^2_F \epsilon^2 E^2 $.
For an average \no energy of 100 MeV, this leads to a mean free path
$l_{n^0}\sim 1.5/\epsilon^2$ so that trapping in the supernova can occur
for $\epsilon \geq 4\times 10^{-3}- 3\times 10^{-2}$. As has been noted
in \cite{babu}, this value of $\epsilon$ is quite adequate to suppress the
contribution of \no to nucleosynthesis.
 
 It should however be noted that while this can evade the energy
constraints, they
will be of no help in eliminating the constraint from considerations of
shock energy and propagation time to be discussed in the last section of
the paper.

\subsection{Galactic $e^+e^-$ population and Early $\gamma$ Ray Flash
  from SN1987A}

As indicated above the observed neutrino signal from SN1987A
  provides a sweeping argument against the \no\, hypothesis.
Here we note that the timing, intensity and character of the ordinary
electromagnetic signals from SN1987A, and from other supernovae,  also 
exclude  a wide range of $\tau$  (or equivalently $B_r$)
values. Moreover,  commulative effects of all past supernovae in the
history of the galaxy also lead to strong constraints.
The key assumption required to enable these constraints is :

A2:
{\it The electromagnetic decay mode $n^0\to e^+e^-\nu_d$ is (one
of) the dominant decays of \no.}

Note that A2 fails to hold if invisible decays like $n^0\to 3\nu$
dominate. This assumption implies that $\tau(n^0\to e^+e^-\nu_d)$ is
also (close to)  the actual $\tau_{n^0}$, the lifetime of \no.

For $\tau\ges 100{\rm s}$ at least $20\%$ of the \no\, emitted will
decay outside the progenitor of SN1987A radius ($\sim 3\times
10^{12}{\rm cm}$) where the average velocity of \no\, was assumed to
be $\beta\sim 0.6$. The total number of decay positrons is then
$N_{e^+}\approx  N(n^0\  {\rm decays\, for} \   r>R)\ \approx 0.3
  N_{n^0}\approx
0.2 f_B N_{\nu_i}\approx 4\times 10^{54}\div 2\times 10^{56}$,
where $N_{\nu_i}\approx 10^{57}$ is the total number of emitted
neutrinos of any given species (say $ \bar{\nu_{\mu}}$). If  the \no\,
production is dominated by 
 $\mu^++n\to p+n^0$ rather than by $\bar{\nu}_{\mu}+N\to N'+n^0$ and
 we use $T\approx 10{\rm  MeV}$ the $f_B$ can be as small as $10^{-3}$.
Following
  \cite {DGN} we note that the decay positrons
   could manifest in two different ways:

(i)
positrons can annihilate on electrons (from decays of different \no
particles as well as ambient) in the vicinity of the supernova progenitor.

(ii)
Positrons that manage to escape into the interstellar medium, slow down and
keep accumulating over the galactic history. A steady rate of
annihilation yielding the monochromatic 512 KeV, $\gamma$-Ray line is
then established. Dar Goodman and Nussinov in a work that preceded the
explosion of SN1987A,  focused on this aspect.  It was found that even 
for  $f_B$ as small as $10^{-3}$, an extremely broad range:
  $\tau_{\nu}\geq 10^4{\rm s}$ \footnote{such a large $\tau$ is 
  required for the particles to traverse red giants envelopes}
 is excluded by the observational limits on the 
  $512 {\rm KeV}$ annihilation line.
The occurrence of SN1987A and the lack of any observations of an early
$\gamma$-Ray pulse, delayed by $R/c\sim 100{\rm s}$ ($R\approx 3\times 
19^{12}\ {\rm cm}$ is the radius of the blue supergiant progenitor of SN198978A),
 with respect to the
neutrino signal, strongly limit electromagnetic decay at $\tau\approx
100{\rm s}$.
Indeed for $\tau\les 100{\rm s}$ most of the \no\, s will decay within a
distance $r\approx R$ from the progenitor. The resulting electron column
density  $ \bar{n_e} R\approx\frac {N_{n^0}}{4\pi R^2} \approx\frac{
    N_{e^+}}{4\pi R^2}=10^{27}\div 10^{30}{\rm cm}^{-2}$,
of electrons or positrons suffices as $ \bar{n_e}
R\sigma_{annihilation}\gg 1$, ensuring almost complete
annihilation. This results in a $\gamma$ fluence
$  \sim \frac{N_{e^+}}{4\pi D^2}\approx 3\times (10^5\div 10^8){\rm cm}^{-2}$

corresponding to energy fluence $(2\div 200){\rm erg\, cm}^{-2}$, for a distance of
$50{\rm kpc}$ to SN1987A.
Such a fluence is $6\div 10$ orders of magnitude larger than the
observational upper bound ! \cite{araa}

Hence, once $N_{e^+}/N_{\nu_i}\ges 10^{-4}$ i. e. once $f_B\geq
10^{-3}$ and $ \Gamma(n^0\to e^+e^-\nu_d)/\Gamma(n^0)\approx  O(1)$
 (namely A2 holds) we can exclude not only $\tau\approx 100{\rm
  s}$ but even $\tau\approx 10{\rm s}$ as well, when the number of \no\, 
decaying
outside the progenitor is suppressed by an additional multiplier of $e^{-R/(c \beta\tau)}\approx
e^{-16}$! (when $\beta\sim 0.6$).

Next, we turn our attention to even shorter lifetimes and show that
also \no\, decays within the progenitor could have dramatic
observational imprints on SN1987A. The lack of these imprints, severely
constraining the allowed $\tau$ values.

\subsection{Shock Energy and Propagation time}

In the standard type-II supernova scenario, the bounce shock from the
collapsed neutron  core propagates outwards and traverses the
progenitor mantle and envelop. In the process, almost all the shock
energy is converted into bulk kinetic energy of the ejected
envelop. When the shock reaches the surface of the star (actually when
it gets to optical depth $\les 1$) the heated up stellar gas emits
radiation in the UV and optical. This marks the beginning of the
optical light curve.  Thus, for SN19897A,  the numerical
hydro-dynamical models \cite{Nomo}, \cite{blin} indeed are capable of reproducing the $3$ hr
time delay between the neutrino pulse and the beginning of the optical
light curve. 
     The propagation time \cite{blin} can be approximated as 

\begin{equation}
T_{prop}=1.4 \left(\frac{R}{R_{\odot}}\right)\left(\frac {M_{ej}}{10
M_{\odot}}\right)^{1/2} \left(\frac{E}{1\times 10^{51}{\rm
erg}}\right)^{-1/2}{\rm hr}
\end{equation}
where  $M_{ej}$, the ejected mass, is the mass exterior to the radius where
the energy $E$ is deposited.

In the present case, one expects to have in addition to the standard
(weakened) bounce shock, a shock resulting from the thermalisation (on
spot essentially) of annihilated $e^+e^-$ resulting from the decay of 
the \no\, particles to $e^+e^-$ that occurs within the mantle or
envelope. This will occur for lifetime of the \no\, particles $\tau\les 
30\ {\ s}$ and  at a mean radius of
$r=v_{n_0} \tau$, with a spread in $r$ which is of order $r$. The
energy in this shock is an order of magnitude larger than that in
the standard bounce shock.
The total amount of heat energy deposited by the  $n^0$ s is actually
their share in the thermal energy of the proto-neutron star core, i.e.
$E_{exp}\sim 3\times 10^{52}{\rm erg}$. This is in contarst with the
small fraction of the neutrino energy that is converted into shock energy.
Thus one expects this shock to be both more vigorous and also
arrive earlier at the stellar surface,

Assuming emission at a temperature of $5\div 10{\rm MeV}\
\beta_{n^0}\approx 0.7$.
 Using the initial stellar structure given by
\cite{blin} which is based on the model of \cite{Nomo},
In the left panel of figure 1 we present the propagation time computed from equation (27)
 as function of $\tau$.

We see that for $\tau\ges 0.03{\rm s}$ the propagation time is
substantially shorter than the observational value of $3 {\rm hr}$. In
the present framework, a propagation time of $3 {\rm hr}$ requires the
progenitor to have a much larger radius which would have rendered it a {\it red}
supergiant while it is known to have been a blue supergiant at the onset
of core collapse.
Also, the fact that the hydrogen-rich envelop would have been pushed out
before the (now weakened) bounce shock from the core collapse could have
reached it, is
expected to substantially reduce the mixing of hydrogen into the iron rich
layers, contrary to the observations.

\begin{figure}[htbp]
 \epsscale{1}
\plottwo{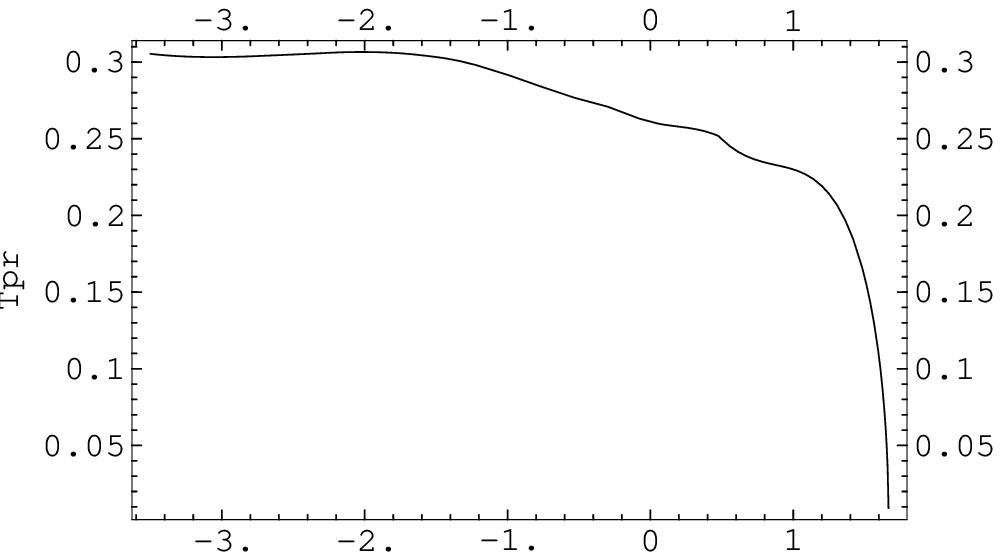}{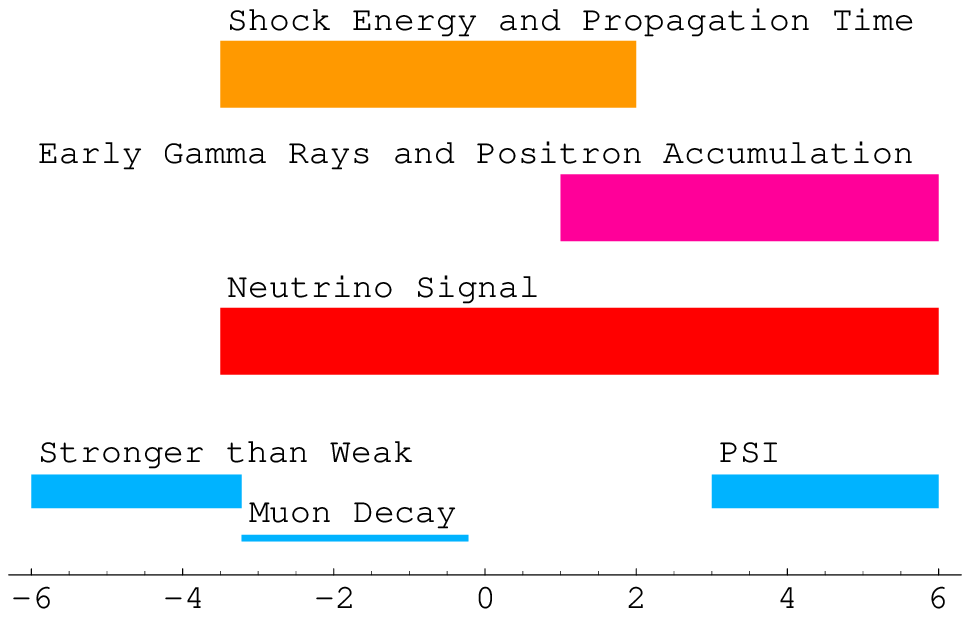}
\centerline{
log $\tau$ [s]}

 \caption{Left: Shock propagation time, in hours,  as function of the
  life-time in seconds. Right: Ranges of life-times excluded by various effects}
 
\end{figure}

\subsection{Summary and Conclusions}

We have seen above that the various considerations pertaining to
SN1987A, and supernovae in general, rule out a large range of possible 
values of $\tau\  (B_r)$. In tyhe right panel of figure 1, we present for each set of
considerations, the range of $\tau$ values which are ruled
out. We also present in the same figure the regions of $\tau$ which
are excluded by the various particle physics considerations
presented in \S 2. Some overlap, so that certain ranges of $\tau$ are
excluded by more 
than one argument. The wide range of the ruled out $\tau$ values
strongly suggests that the \no\, hypothesis is not viable.

We dedicate this paper to our colleague Arnon Dar on the
occasion of his 60th birthday. We would like to thank  F. T. Avignoni, 
C. Rosenfeld, and S. Mishra for early discussions on the KARMEN
anomaly and Y. Alster, D. Asheri, A. Kerman, and E. Piasetzky for
discussions on the neutron hypothesis.
The work of all three authors has been supported by the
Israel--US bi-national fund, grant  94-314. The work of I.G. and
S.N. also by the Israel national Science Foundation, grant 561/99
and the work R. N. M. is supported by the NSF grant no. PHY-9802551.

\end{document}